\documentclass[%
 reprint,
superscriptaddress,
 amsmath,amssymb,
 aps,
]{revtex4-2}

\usepackage{graphicx}
\usepackage{dcolumn}
\usepackage{bm}

\usepackage{physics}
\usepackage{upgreek}
\usepackage{gensymb}
\usepackage{units}
\usepackage{float}
\usepackage{amsmath}
\usepackage[colorlinks=true,linkcolor=blue,urlcolor=blue,citecolor=blue]{hyperref}

\usepackage{float}
\usepackage[commentmarkup=uwave,deletedmarkup=sout]{changes}
\definechangesauthor[name=Malte Grosche, color=blue]{fmg}
\begin{document}
\title{The Fermi surface of RuO$_2$ measured by quantum oscillations}

\author{Zheyu Wu}
\author{Mengmeng Long}
\author{Hanyi Chen}
\affiliation{Cavendish Laboratory, University of Cambridge,\\
 JJ Thomson Avenue, Cambridge, CB3 0HE, United Kingdom}

\author{Shubhankar Paul}
\affiliation{Toyota Riken–Kyoto University Research Center (TRiKUC), Kyoto 606-8501, Japan}
\affiliation{Department of Physics, Indian Institute of Technology Kanpur, Kanpur 208016, India}

\author{Hisakazu Matsuki}
\affiliation{Toyota Riken–Kyoto University Research Center (TRiKUC), Kyoto 606-8501, Japan}

\author{Oleksandr Zheliuk}
\author{Uli Zeitler}
\affiliation{High Field Magnet Laboratory (HFML-EMFL), Radboud University,\\
Toernooiveld 7, 6525 ED Nijmegen, The Netherlands}

\author{Gang Li}
\author{Rui Zhou}
\affiliation{Beijing National Laboratory for Condensed Matter Physics, Institute of Physics, Chinese Academy of Sciences, Beijing 100190, China}
\affiliation{School of Physical Sciences, University of Chinese Academy of Sciences, Beijing 100190, China}

\author{Zengwei~Zhu}
\affiliation{Wuhan National High Magnetic Field Center, Wuhan 430074, China}


\author{Dave~Graf}
\affiliation{National High Magnetic Field Laboratory, Tallahassee, Florida, 32310, USA}

\author{Theodore~I.~Weinberger}
\author{F. Malte Grosche}
\affiliation{Cavendish Laboratory, University of Cambridge,\\
 JJ Thomson Avenue, Cambridge, CB3 0HE, United Kingdom}

\author{Yoshiteru Maeno}
\email{maeno.yoshiteru.b04@kyoto-u.jp}
\affiliation{Toyota Riken–Kyoto University Research Center (TRiKUC), Kyoto 606-8501, Japan}

\author{Alexander G. Eaton}
 \email{alex.eaton@phy.cam.ac.uk}
\affiliation{Cavendish Laboratory, University of Cambridge,\\
 JJ Thomson Avenue, Cambridge, CB3 0HE, United Kingdom}
 
\date{\today}

\begin{abstract}
\noindent
The metallic oxide RuO$_2$ has emerged as a promising altermagnet candidate, owing to reports of this material hosting antiferromagnetic ordering accompanied by a spin-split electronic band structure characteristic of time-reversal symmetry-breaking. However, recent studies have robustly questioned this scenario. Here we map the Fermi surface of pristine single-crystalline RuO$_2$. By measuring magnetic quantum oscillations of a bulk thermodynamic property, our study resolves the electronic structure present in the bulk of RuO$_2$. Several Fermi sheets are discerned, with a range of effective quasiparticle masses up to five times that of the bare electron mass. We compare our measurements with the predictions for altermagnetic and nonmagnetic Fermi surfaces deduced from density functional theory calculations. The quantum oscillatory frequency spectra correspond very poorly to the profile expected for the case of altermagnetism; by contrast, they correspond well to the nonmagnetic scenario. Our findings place significant constraints on the bulk magnetic properties of RuO$_2$, and strongly suggest that this material is a paramagnet.


\end{abstract}

\maketitle 
\section{Introduction}
The recently proposed  phenomenon of an altermagnetic (AM) phase of matter -- characterized by a compensated collinear magnetic structure with a spin polarization that alternates through both the crystal structure in real space and the electronic band structure in reciprocal space -- has attracted widespread attention~\cite{BeyondPRX.12.031042,EMERGINGPRX.12.040501,KyoPRB.99.184432,naka2019spin,savitsky2024science,hayami2019momentum,MazinEditPRX.12.040002,smejkal2020sciadv,fender2025altermagnetism,song2025altermag-rev}. In addition to being of great fundamental interest -- constituting a novel, distinct class of magnetic order -- there are numerous properties of AM materials that make them highly desirable for technological applications~\cite{bai2024altermagreview,khalili2024annrev}. For example, their time-reversal symmetry-broken electronic band structures may, as per ferromagnetic materials, manifest spin-polarized currents accompanied by the anomalous Hall effect~\cite{naka2019spin,KyoPRB.99.184432}. Furthermore, attempts to miniaturize ferromagnetic components within integrated circuits often suffer deleterious effects from stray fields causing interference between neighboring elements. By contrast, the lack of a net magnetization \textbf{M} for AM components could enable them to be spatially positioned much closer together in next-generation spintronic circuitry.

Several materials have been theoretically predicted to exhibit AM ordering~\cite{hayami2019momentum,smejkal2020sciadv}. Amongst these, the rutile compound RuO$_2$ quickly emerged as an especially promising candidate~\cite{KyoPRB.99.184432,smejkal2020sciadv}. Antiferromagnetic ordering in this material was inferred from measurements of neutron diffraction~\cite{BerlijnPRL17} and x-ray scattering~\cite{ZhuPRL19}, with these experiments interpreted to indicate the presence of compensated magnetic moments aligned along the rutile $c$-axis. Several predicted AM properties were reported from subsequent experimental studies, including observations of the anomalous Hall effect~\cite{feng2022anomalous}, spin-current generation~\cite{bose2022tilted,BaiPRL22,KarubePRL22,guo2024direct} along with magnetic circular dichroism indicative of time-reversal symmetry-breaking in the band structure~\cite{fedchenko2024TRSBSciAdv}.

However, the widely held interpretation of RuO$_2$ being an archetypical altermagnet has recently been strongly challenged. Firstly, muon spin rotation measurements were unable to detect any signatures of magnetic ordering down to a detection limit of $< 1 \times 10^{-3} \mu_{\text{B}}$ per Ru site, where $\mu_{\text{B}}$ is the Bohr magneton~\cite{HiraishiPRL24,kessler2024absence}. Secondly, while some photoemission experiments were interpreted to have resolved the $d$-wave spin pattern characteristic of AM ordering~\cite{fedchenko2024TRSBSciAdv,lin2024observationgiantspinsplitting}, subsequent photoemission studies reported no evidence of AM spin-splitting in the band structure~\cite{Liu-PRL.133.176401,sato-arpes-arxiv}. Instead, heavy surface states were identified, of a possible topological origin. A Rashba-like spin-splitting at the surface was also reported~\cite{Liu-PRL.133.176401}, which may explain the observation of magnetic circular dichroism~\cite{fedchenko2024TRSBSciAdv}. Given this controversy between surface-sensitive studies, a clear unambiguous determination of the bulk electronic band structure of RuO$_2$ is therefore urgently required to shed light on the intrinsic electronic and magnetic properties of this material.

Here we report de Haas-van Alphen (dHvA) and Shubnikov-de Haas (SdH) effect measurements of quantum oscillations (QOs)~\cite{Shoenberg1984} in the magnetic torque and contactless resistivity of pristine quality single crystalline RuO$_2$. QO measurements have been proposed as an ideal diagnostic tool for probing altermagnet candidates, due to this technique's high fidelity for resolving the spin-splitting of electronic bands such materials must necessarily possess~\cite{li2024diagnosingaltermagneticphasesquantum}. We measured the evolution in QO frequency upon rotating the orientation of applied magnetic field \textbf{H} between the [001]-[100] and [100]-[010] crystallographic axes, to probe the geometrical and topological properties of the RuO$_2$ Fermi surface. We compare our experimental observations with theoretical calculations, and find our measurements to be very well described by RuO$_2$ being a paramagnetic metal possessing no magnetic ordering.


\section{Experimental techniques}

Single crystals of RuO$_2$  used in this study were grown by the vapor-transport method in flowing oxygen. The details of the growth condition are described in ref.~\cite{sato-arpes-arxiv}, with most of the crystals used in this study coming from the same crystal growth batch as used in that photoemission study~\cite{sato-arpes-arxiv}. 
Temperature dependence of the electrical resistivity shows metallic behavior with a residual resistivity of 0.1 $\upmu \Omega$cm for current along the [001] direction, with a residual resistivity ratio of 400, indicative of pristine sample quality.

We performed measurements in four separate magnet systems. Capacitive torque magnetometry experiments were undertaken in steady magnetic fields up to 31 T in a resistive magnet at the High Field Magnet Laboratory (HFML-EMFL), Nijmegen, The Netherlands and up to 30 T in a superconducting magnet at the Synergetic Extreme Condition User Facility (SECUF), Beijing, China. The HFML-EMFL magnet was fitted with a $^3$He sample environment attaining a base temperature of 0.4~K, while in SECUF we utilized a dilution refrigerator enabling temperatures as low as 60~mK. We also measured the contactless resistivity of RuO$_2$ by the tunnel diode oscillator (TDO) and proximity detector oscillator (PDO) techniques~\cite{RevSciIns_TDO,PDO_Altarawneh}. TDO measurements were performed at SECUF and at the National High Magnetic Field Laboratory (NHMFL), Florida, USA, in a resistive magnet attaining 41.5~T at a base temperature of 0.4~K. PDO measurements were taken at the Wuhan National High Magnetic Field Center (WNHMFC), Wuhan, China. Our PDO experiments were performed in a pulsed magnet reaching a maximum field strength of 57~T, at a base temperature of 0.7~K.

For torque magnetometry measurements, samples were aligned by Laue diffractometry and affixed to a BeCu cantilever using a low-temperature adhesive varnish. This technique constitutes a bulk-sensitive thermodynamic measurement that provides a highly sensitive probe of the anisotropy in a material’s magnetic susceptibility tensor. Due to its high fidelity, torque magnetometry is often deployed for dHvA effect measurements in high magnetic fields~\cite{bergemann1999torqueQOs,sebastian2008multi,MatsuyamaPRB23,Eaton2024}. In HFML-EMFL the magnetic torque $\boldsymbol{\tau} = \mu_0 (\textbf{M} \cross \textbf{H})$ was measured capacitively by a General Radio analogue capacitance bridge, and calibrated to absolute units using an Andeen-Hagerling (AH) digital capacitance bridge. At SECUF, we measured $\tau$ directly with an AH bridge. In both sets of torque experiments, samples were rotated in-situ, with the orientation calibrated by a Hall sensor. For contactless resistivity measurements, samples were also aligned by Laue diffractometry, and affixed to hand-wound copper coils by low-temperature adhesive varnish. TDO and PDO measurements were obtained utilizing a similar methodology to that outlined in ref.~\cite{theo2024}. At WHMFC, samples were also rotated in-situ, with the orientation determined by a pick-up coil.


\begin{figure*}[t!]
    \includegraphics[width=1\linewidth]{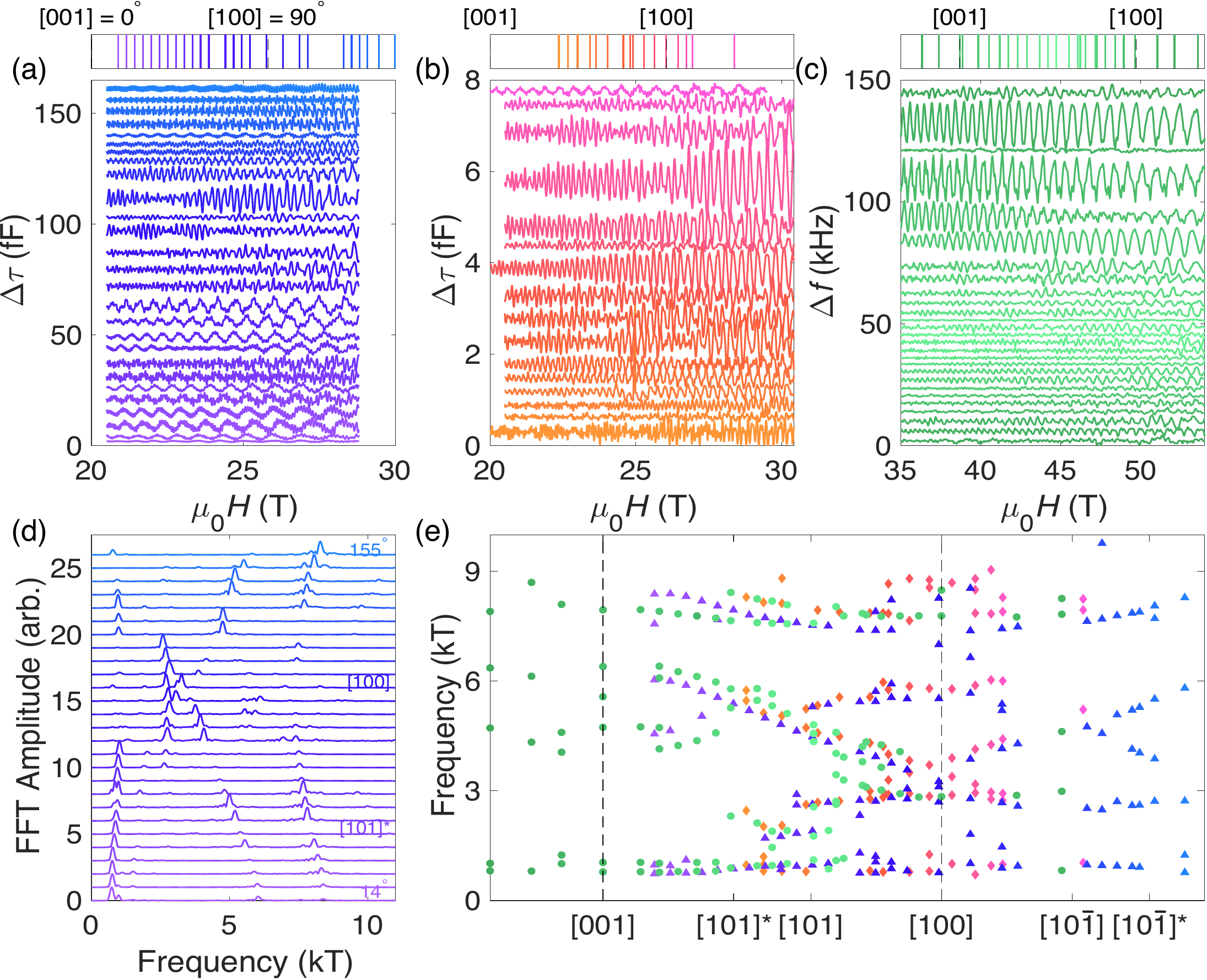}
    \caption{Angle-dependent quantum oscillation measurements in RuO$_2$. Background-subtracted magnetic torque ($\Delta \tau$) measured at incremental angles in the [001]—[100] rotation plane recorded at (a) SECUF and (b) HFML-EMFL. (c) Quantum oscillations in the contactless resistivity measured at WNHMFC, also in the [001]—[100] plane. (d) Fast Fourier transforms (FFTs) of the data from panel \textit{a}. These spectra have been renormalized and offset for ease of presentation. (e) Angular distribution of quantum oscillatory frequency spectra, determined from performing FFTs for the data in panels \textit{a-c}, with symbols plotted in the same colors as the waveforms. Only fundamental components are included. Good correspondence is observed between the three measurements. Multiple frequency branches are resolved, indicating a complex multi-sheet Fermiology.}
    \label{fig:c-awiggles}
\end{figure*}

\begin{figure*}[t!]
    \includegraphics[width=1\linewidth]{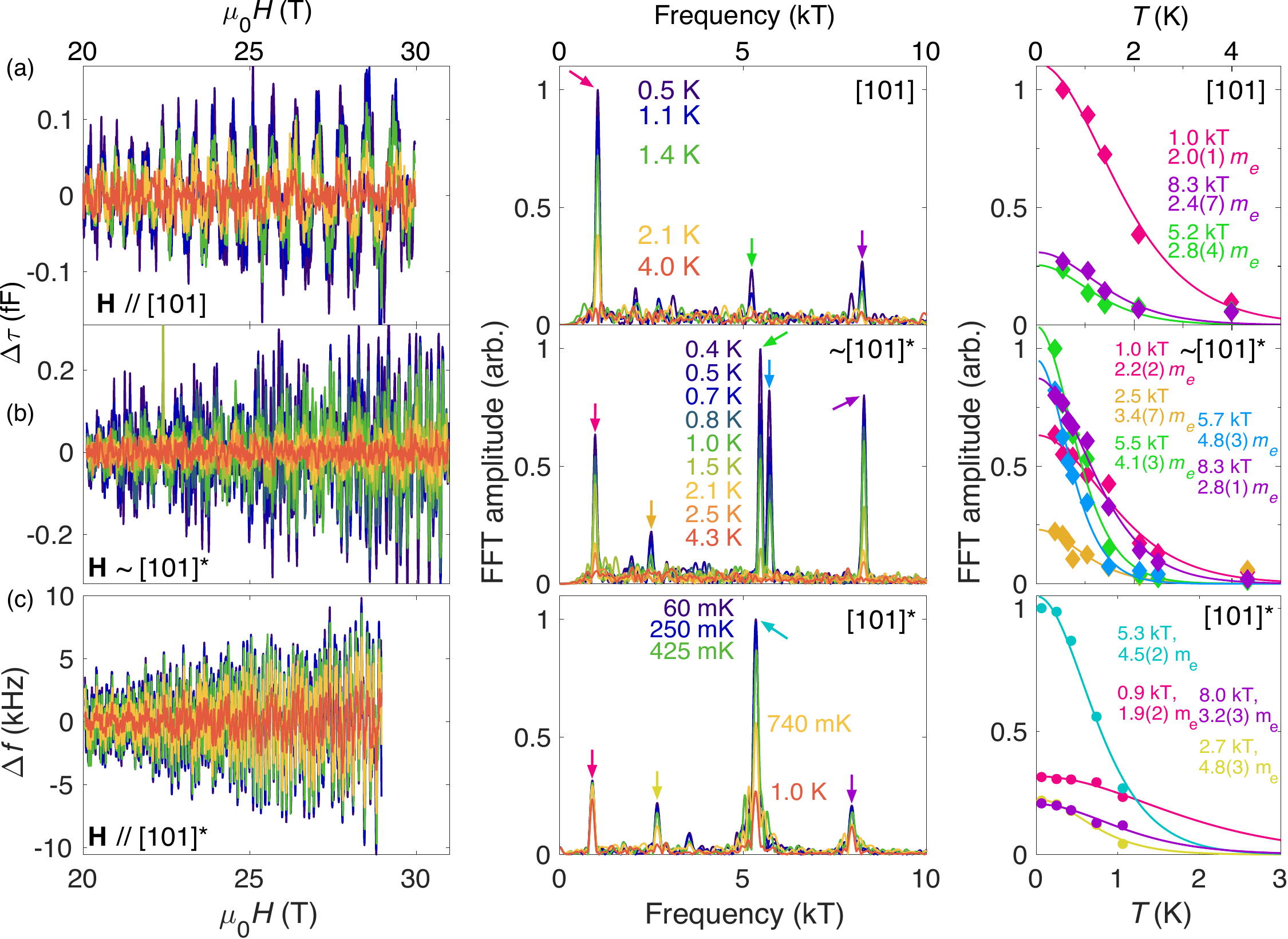}
    \caption{Temperature evolution of (left) quantum oscillatory waveforms in RuO$_2$, (centre) their corresponding FFT spectra and (right) Lifshitz-Kosevich fits to the $T$-dependence of FFT amplitudes for (a) torque measurements with \textbf{H} $\parallel$ [101], (b) torque measurements with \textbf{H} tilted three degrees away from [101]* towards [100] and (c) TDO measurements with \textbf{H} $\parallel$ [101]*. Multiple frequency branches are observed ranging from 900~T to 8.3~kT, with a spread of effective quasiparticle masses up to 4.8(3)~$m_e$.}
    \label{fig:temps}
\end{figure*}

\section{Results}
Fig.~\ref{fig:c-awiggles} presents QOs from our SECUF, HFML-EMFL and WNHMFC experiments, for rotations of the orientation of \textbf{H} through the [001]-[100] rotation plane. Fig.~\ref{fig:c-awiggles}e collates the results of these experiments by plotting the oscillatory frequencies versus rotation angle, where the frequencies have been computed by fast Fourier transforms (FFTs) performed on the QO traces at each angle. Good correspondence between the three datasets is observed, with several frequency branches clearly resolved.

In Appendix~\ref{appx_harmonics} we examine the harmonic content of the quantum oscillatory spectra. The pristine quality of single crystal specimens measured in this study is underlined by the observation of an eighth harmonic component at low temperatures and high magnetic fields, at a frequency of 71.5~kT, indicating a mean free path $\gtrsim$ 0.5 $\upmu$m. This tallies with the residual resistivity of 0.1~$\upmu \Omega$cm, which by simple Drude considerations implies a mean free path of $\approx 0.9$~$\upmu$m. Throughout the present section we focus solely on the fundamental QO frequency components. 

In Fig.~\ref{fig:temps} we plot the evolution in temperature $T$ of the oscillatory waveforms for torque measured with \textbf{H} aligned along the [101] direction and close to the [101]* direction, and for TDO with \textbf{H}~$\parallel$~[101]*. (See Appendix~\ref{appx_orient} for a discussion of the crystal orientation.) Numerous frequency components with a range of effective cyclotron masses are resolved. Good correspondence is observed between the frequency spectra and effective masses resolved in both our magnetic torque and contactless resistivity experiments. As $\boldsymbol{\tau}$ is a bulk thermodynamic property of the material, this close correspondence indicates that our contactless resistivity measurements are sensitive to the bulk band structure of RuO$_2$.

\begin{figure*}[t!]
    \includegraphics[width=1\linewidth]{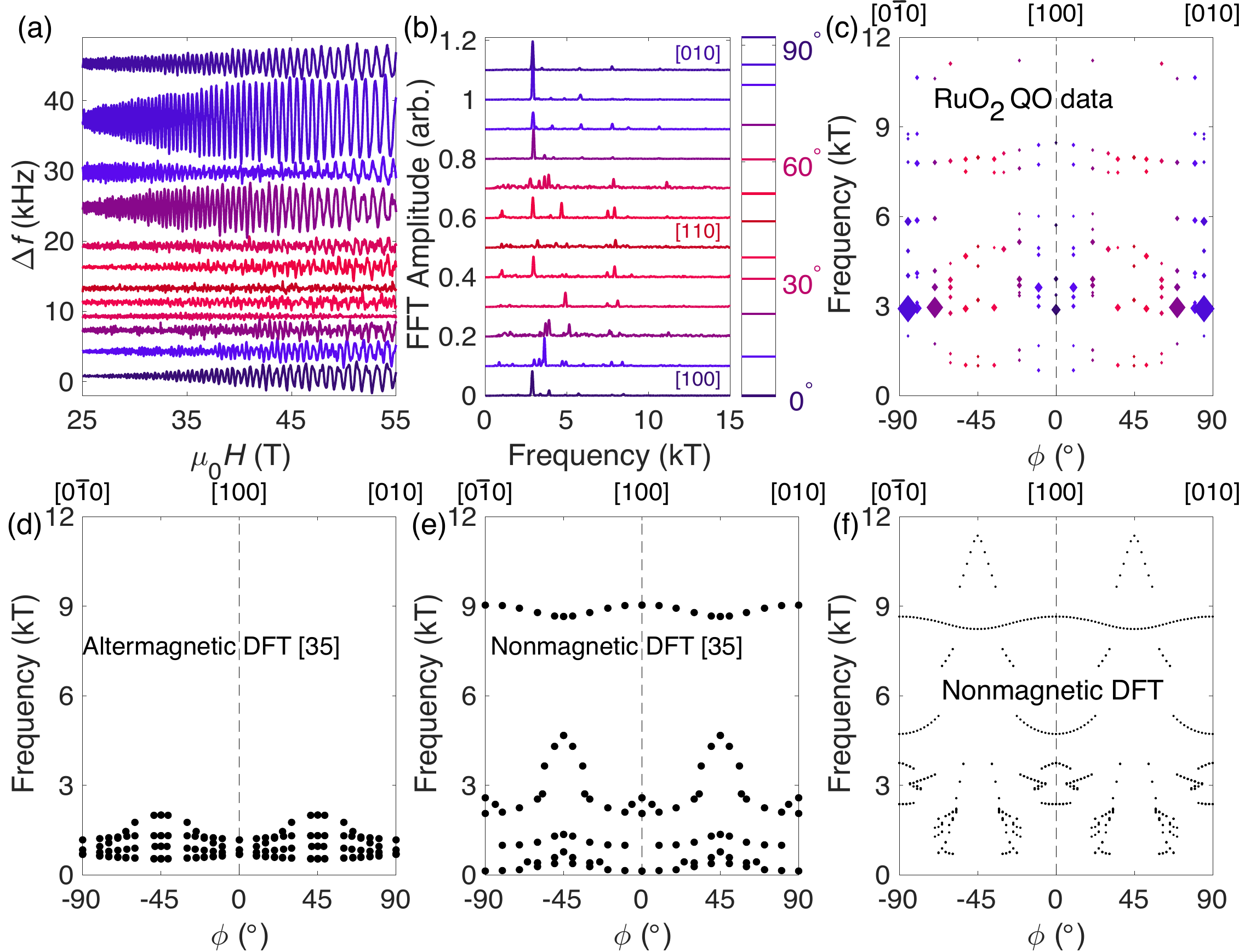}
    \caption{Quantum oscillations in the [100]-[010] rotation plane. (a) SdH effect measurements of RuO$_2$ measured at WNHMFC with (b) the corresponding FFT spectra. (c) QO frequency versus rotation angle $\phi$, where $\phi = 0\degree$ corresponds to \textbf{H} $\parallel$ [100] while $\phi = 90\degree$ indicates \textbf{H} $\parallel$ [010]. The size of each point corresponds to the amplitude of that frequency component's FFT peak. Negative angles are plotted for illustrative purposes, reflected through $\phi = 0\degree$. A strong dependence of quantum oscillatory amplitude on magnetic field orientation is observed. (d) The predicted QO frequency versus angle profile for RuO$_2$ possessing AM ordering reported by ref.~\cite{RuO2QODFT_PRB24} and (e) the expected profile in the absence of magnetic ordering. (f) The predicted angular QO frequency evolution from our DFT calculations for the nonmagnetic Fermi surface of RuO$_2$. A complex structure of several frequency components is expected, similar to the experimental observation in panel \textit{c}.}
    \label{fig:a-bwiggles}
\end{figure*}

In Fig.~\ref{fig:a-bwiggles} we probe the Fermi surface by rotating \textbf{H} between the [100] and [010] axes. Due to the tetragonal crystal symmetry, the frequency spectra should be degenerate for \textbf{H} $\parallel$ [100] and $\parallel$ [010]. However, the frequency profile in Fig.~\ref{fig:a-bwiggles}c shows subtle differences between these two orientations. This is likely due to a slight misalignment of $\sim 1\degree$ causing a small but discernible spread of frequencies due to the presence of numerous extremal orbits about a complex Fermi surface structure.

Prior density functional theory (DFT) calculations~\cite{RuO2QODFT_PRB24} predicted that the observed QO frequencies should yield starkly different angular profiles depending on whether RuO$_2$ is altermagnetic or paramagnetic. Fig.~\ref{fig:a-bwiggles} compares the angular dependence of QO frequencies observed by experiment with the DFT predictions for nonmagnetic and AM ordering reported in ref.~\cite{RuO2QODFT_PRB24}, along with the results of our own DFT calculations (see Appendix~\ref{apx_dft} for details). While the predicted AM Fermi surface from ref.~\cite{RuO2QODFT_PRB24} would not be expected to yield any frequency components $> 3$~kT in this rotation plane, the nonmagnetic Fermi surfaces from that study and from our calculations should both manifest a range of QO frequencies including a prominent branch at $\approx 9$~kT. This prediction matches well with what we observe experimentally. In Fig.~\ref{fig:a-bwiggles}c we resolve numerous frequency components ranging from 850~T to 11~kT. 

\begin{figure*}[t!]
    \includegraphics[width=\linewidth]{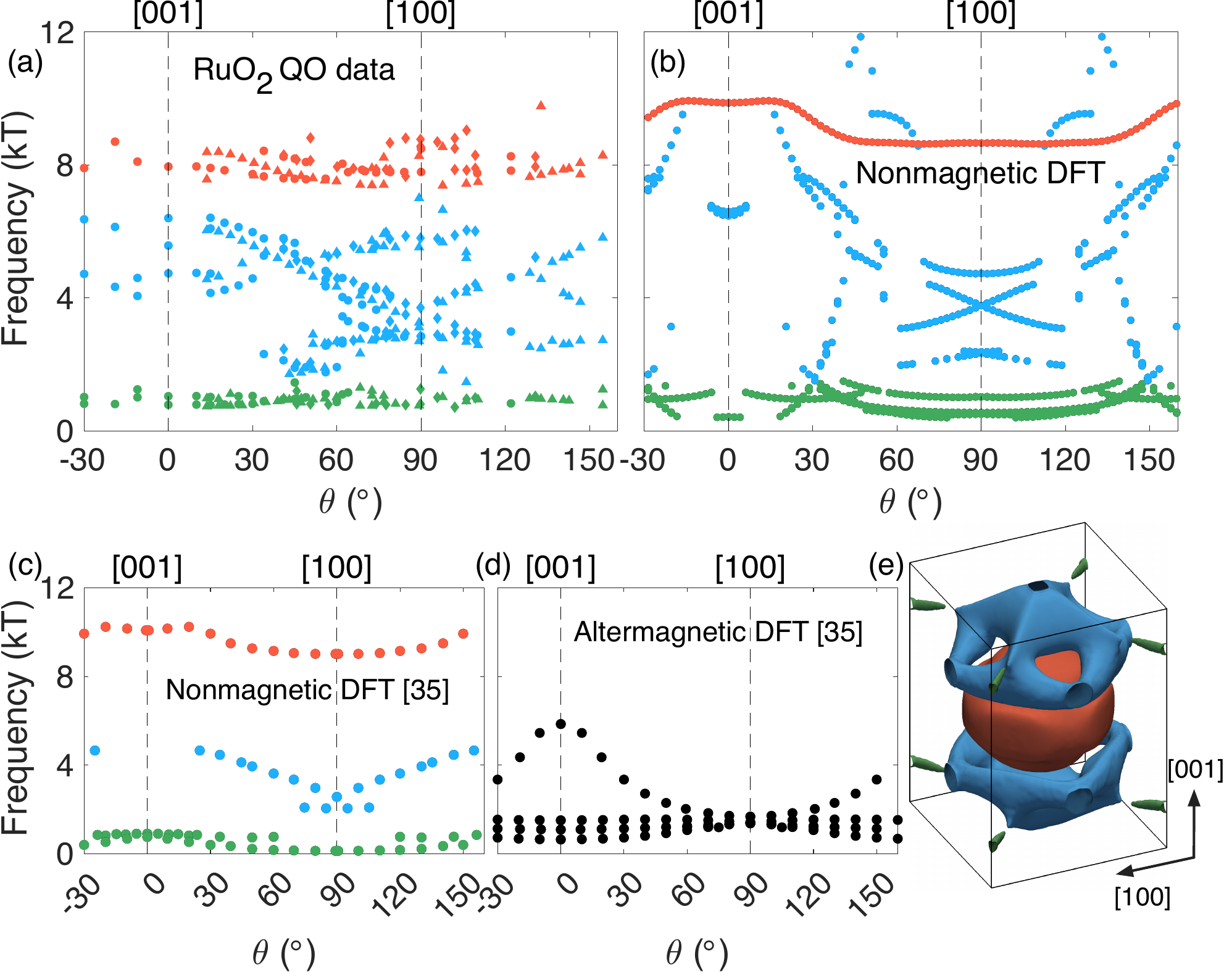}
    \caption{Angular evolution of quantum oscillatory frequencies of RuO$_2$ in the [001]-[100] rotation plane. (a) Fundamental QO frequencies versus rotation angle $\theta$ reproduced from Fig.~\ref{fig:c-awiggles}. Here $\theta = 0\degree$ corresponds to \textbf{H} $\parallel$ [001] and $\theta = 90\degree$ to \textbf{H} $\parallel$ [100]. (b) Simulated QO frequencies versus $\theta$ computed from our DFT calculations for a nonmagnetic Fermi surface. (c) The predicted angular frequency distribution in the absence of magnetism from ref.~\cite{RuO2QODFT_PRB24} and (d) the alternative prediction for the case of AM ordering. (e) Rendering of the nonmagnetic RuO$_2$ Fermi surface from our DFT calculations. The red, blue and green coloring throughout this figure is indicative of the three separate Fermi sheets.}
    \label{fig:FSandangles}
\end{figure*}

In Fig.~\ref{fig:FSandangles} we collate the experimentally resolved QO frequency spectra in the [001]-[100] rotation plane from Fig.~\ref{fig:c-awiggles} and compare with the expectations from DFT calculations. Numerous features of the nonmagnetic Fermi surface predictions match well with our experimental observations. These include a largely isotropic high frequency branch at around 8-9~kT, a low frequency branch around 1~kT, and a more complex intermediate branch that exhibits greater variation of frequency with angle than the other two branches. By contrast, the predicted AM angular frequency profile corresponds very poorly to our measurements.



\section{Discussion}
The presence of magnetic ordering in a material may introduce new periodic potentials and induce hybridization between localized and itinerant states, thereby opening energy gaps and shifting the electronic bands~\cite{peierls1955quantum}. This typically leads to a significant reconstruction of the Fermi surface compared to the nonmagnetic scenario. The angular dependence of the quantum oscillatory frequency spectra that we observe in our experiments (Figs.~\ref{fig:a-bwiggles}~\&~\ref{fig:FSandangles}) therefore sets strong constraints on the possible magnetic properties present within the bulk of RuO$_2$.

If AM ordering were to be manifested throughout the bulk of RuO$_2$, a markedly different Fermi surface geometry and topology has been expected~\cite{EMERGINGPRX.12.040501,RuO2QODFT_PRB24} compared to the nonmagnetic case. The DFT calculations of both the present study and ref.~\cite{RuO2QODFT_PRB24} for the scenario of no magnetic ordering in RuO$_2$ capture the main features of the quantum oscillatory frequency spectra resolved by experiment. Our data are well described by three Fermi surface sheets (Fig.~\ref{fig:FSandangles}). These include two largely isotropic, almost spherical pockets, one yielding QO frequencies $\approx$~8-9~kT, and another much smaller one at $\approx$~1~kT. The other Fermi surface sheet (colored blue in Fig.~\ref{fig:FSandangles}e) possesses a much more complex geometry, consisting of interconnected body and neck sections with numerous extremal orbital areas contributing to the observed QO frequency spectra. We note that this interpretation of three quite distinct Fermi sheets is broadly consistent with much earlier work on RuO$_2$~\cite{marcus1968measurement,slivka1968azbel,Graebner76,Mattheiss76,yavorsky1996ab}, in which this material was also considered to be nonmagnetic.

Recent photoemission experiments have discerned the presence of electronic states that appear to be confined to the surface of RuO$_2$ with low dispersion characteristic of low Fermi velocities~\cite{Liu-PRL.133.176401,sato-arpes-arxiv}. However, we do not resolve any signature of these states in either our dHvA or SdH measurements. That torque magnetometry is not sensitive to these surface states is unsurprising, since this technique probes a bulk thermodynamic property. Consider for example a cubic sample geometry -- the ratio of surface to bulk unit cells in such a system is 1:10$^6$, therefore the contributory oscillatory amplitude from any surface component may be expected to be orders of magnitude smaller than that from the bulk~\cite{hartstein2018fermi}. By contrast, contactless resistivity measurements by the TDO and PDO techniques are much more sensitive to surface effects, as they operate at MHz frequencies for which the skin depth of a high conductivity metal such as RuO$_2$ is very short. However, the clearest photoemission spectra of the surface states appear to resolve a largely circular pocket that encloses an area (in reciprocal space) of $\approx 1$~\AA$^{-2}$~\cite{Liu-PRL.133.176401,sato-arpes-arxiv}, which for \textbf{H}~$\parallel$ [110]$^*$ would yield QOs with a frequency $\approx$~10~kT. At $\mu_0 H =$~30~T, the corresponding cyclotron orbit of such a pocket would possess a diameter (in real space) of $\approx$~200~nm, necessitating a long mean free path in the surface layer. While the bulk of our samples indeed possesses such a long mean free path (see Appendix~\ref{appx_harmonics}), if the surface states are topological as has been proposed~\cite{sato-arpes-arxiv} then they will be confined to a very small volumetric fraction of the sample right at the surface. It would be reasonable to assume that the mean free path at the surface is less than in the bulk, which would in turn explain why we did not appear to resolve any signature of these surface states. Additionally, we note that some of the surface sheets appear to be quasi-1D, which in a magnetic field would yield open orbits that do not produce QOs.

The pronounced disagreement between the expected geometry of the AM Fermi surface of RuO$_2$ and our experimental observations corroborates the growing body of evidence in favor of RuO$_2$ not being magnetically ordered~\cite{HiraishiPRL24,kessler2024absence,Liu-PRL.133.176401,sato-arpes-arxiv,WenzelPRB25,kiefer2025jpcm}. Reconciling a nonmagnetic ground state with phenomena including the anomalous Hall effect, spin-current generation and magnetic circular dichroism~\cite{feng2022anomalous,bose2022tilted,BaiPRL22,KarubePRL22,guo2024direct,fedchenko2024TRSBSciAdv} therefore presents an outstanding challenge. The anomalous character of the electronic surface states resolved by photoemission spectroscopy -- reported to display a Rashba-like spin-splitting~\cite{Liu-PRL.133.176401} with non-trivial topology~\cite{sato-arpes-arxiv} -- may be the cause of several of these observations. It has also been proposed that RuO$_2$ thin films may exhibit a different magnetic character compared to single crystals~\cite{brahimi2024confinementinducedaltermangetismruo2films}, and that the bulk magnetic properties may be acutely sensitive to the density of ruthenium vacancies~\cite{SmolyanyukPRB24-vacancies}. We note that the observation of very high frequency QO components, with multiple harmonics up to 71.5~kT, implies a long mean free path and thus a very low density of vacancies in our measured crystals.

In conclusion, we performed a quantum oscillation study of the metallic altermagnet candidate RuO$_2$. We resolved a Fermi surface geometry that matches well with the predicted electronic structure in the absence of magnetism; whereas, the expected angular dependence of quantum oscillatory frequency spectra for the case of altermagnetic ordering disagrees markedly with our experimental observations. Our findings strongly support the scenario that the magnetic ground state of RuO$_2$ is paramagnetic in character.

%


\vspace{-0mm}
\begin{acknowledgments}\vspace{-5mm}
We gratefully acknowledge stimulating discussions with A. Agarwal, D. Calugaru, D. Chichinadze, A. Coldea, B. Ramshaw, T. Sato, D. Shaffer, R.-J. Slager and S. Souma. This project was supported by the EPSRC of the UK (grant no. EP/X011992/1). We acknowledge the support of HFML-RU/NWO-I, member of the European Magnetic Field Laboratory (EMFL) and the Engineering and Physical Sciences Research Council (EPSRC, UK) via its membership to the EMFL (grant EP/N01085X/1). A portion of this work was carried out at the Synergetic Extreme Condition User Facility (SECUF, \href{https://cstr.cn/31123.02.SECUF}{https://cstr.cn/31123.02.SECUF}). This work was supported by JSPS KAKENHI grant no. JP22H01168. A portion of this work was performed at the National High Magnetic Field Laboratory, which is supported by National Science Foundation Cooperative Agreement No. DMR-2128556 and the State of Florida. S.P. acknowledges support of the JST Sakura Science Program. T.I.W. and A.G.E. acknowledge support from QuantEmX grants from ICAM and the Gordon and Betty Moore Foundation through Grant GBMF9616 and from the US National Science Foundation (NSF) Grant Number 2201516 under the Accelnet program of Office of International Science and Engineering (OISE). A.G.E. acknowledges support from Sidney Sussex College (University of Cambridge).
\end{acknowledgments}

\appendix

\section{Harmonic content of quantum oscillatory frequency spectra}
\label{appx_harmonics}

\normalsize
\vspace{5mm}\noindent

\begin{table}[h!]
\caption{List of harmonic components and their corresponding frequencies and effective masses from the data presented in Fig.~\ref{fig:tally} for \textbf{H} $\parallel$ [001]. Effective masses are only quoted for branches with sufficient amplitude to be well resolved at elevated temperatures. The $n^{\text{th}}$ harmonic is easily identifiable as possessing $n$ times the frequency and mass of the corresponding fundamental orbit~\cite{Shoenberg1984}.}
\begin{tabular}{c|c|c}
         Harmonic index & Frequency (kT) & Effective mass ($m_e$) \\ \hline
$\alpha$  & 0.82     & 2.3(3)               \\
$2\alpha$ & 1.63     & 5(1)               \\ \hline
$\beta$   & 6.46     & 6.6(5)               \\
$2\beta$  & 12.9    & 11(1)               \\
$3\beta$  & 19.4     & -                  \\ \hline
$\gamma$  & 8.95     & 2.5(2)               \\
$2\gamma$ & 17.9    & 5.3(1)               \\
$3\gamma$ & 26.8    & 7.8(3)               \\
$4\gamma$ & 35.7    & 9.8(3)               \\
$5\gamma$ & 44.7    & 12(1)               \\
$6\gamma$ & 53.6    & 13(2)               \\
$7\gamma$ & 62.6    & -                  \\
$8\gamma$ & 71.5    & -                 
\end{tabular}
\label{table1}
\end{table}

\begin{figure*}[h]
    \includegraphics[width=.8\linewidth]{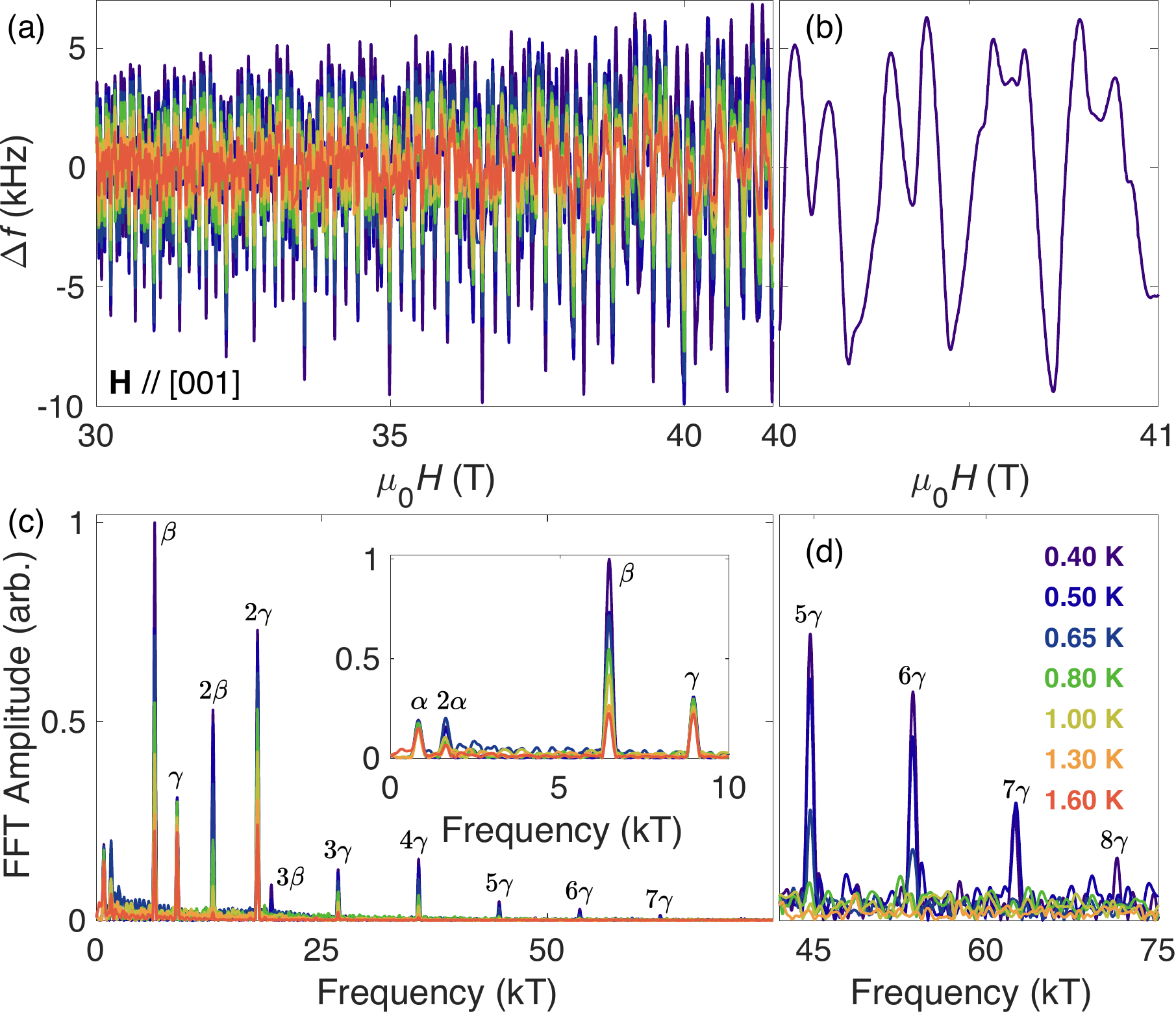}
    \caption{High harmonic frequency components. (a) QOs in the contactless resistivity of RuO$_2$ measured by the TDO technique at NHMFL. (b) Zoom in between 40~T and 41~T of the base temperature sweep in panel \textit{a}, with several fast frequency components visible. (c) FFT of the data in \textit{a}, with numerous harmonics identified. The inset shows the 0-10~kT frequency range, in which the three fundamental frequencies $\alpha$, $\beta$ and $\gamma$ are labeled. (d) Zoom in of the highest frequency Fourier spectra, in which the eighth harmonic of the $\gamma$ branch is resolvable at low $T$.}
    \label{fig:tally}
\end{figure*}

In Fig.~\ref{fig:tally} we present contactless conductivity data for an RuO$_2$ single crystal measured by the TDO method at NHMFL in high magnetic field strengths up to 41.5~T, with the field oriented along the [001] direction. Numerous harmonic components are visible in the Fourier spectra, which are tabulated in Table~\ref{table1}.

A QO frequency $f$ is related to an extremal Fermi surface cross-sectional area (in reciprocal space). This area corresponds to a quasiparticle orbit, in real space, which for a simple circular geometry would possess a radius $r = \sqrt{\frac{2 \hslash f}{e (\mu_0 H)^2}}$ where $e$ is the elementary charge and $\hslash$ the reduced Planck constant~\cite{Shoenberg1984,onsager_rel}. Samples with long mean free paths $\lambda$ are therefore required in order to resolve high frequency contributions, as QOs of frequency $f$ will only be manifested if $2r \lesssim \lambda$. Access to high magnetic fields is valuable, as increasing $H$ reduces $r$. The observation of the eighth harmonic of the $\gamma$ branch, with $f$~=~71.5~T resolved for $\mu_0 H \gtrapprox$~36~T, therefore indicates a long mean free path of $\lambda \gtrsim$~0.5~$\upmu$m.


\clearpage

\section{Crystallographic orientation}
\label{appx_orient}
\normalsize
\vspace{5mm}\noindent 

\begin{figure}[h!]
    \includegraphics[width=1\linewidth]{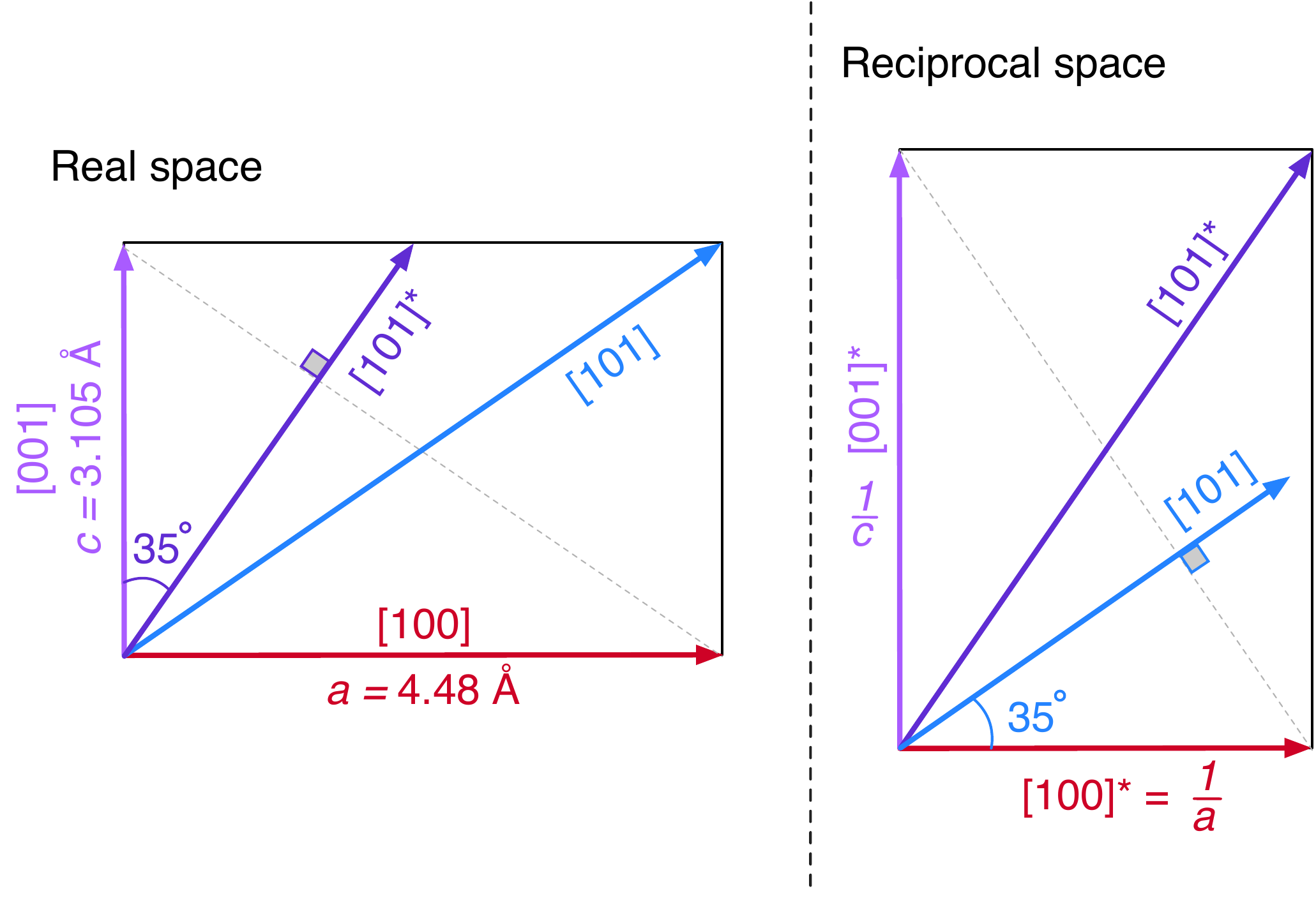}
    \caption{2D representation of the crystal structure of RuO$_2$, with a diagrammatic illustration of the [101] and [101]* directions that are separated by 20$\degree$ in the $c-a$ plane. In real space, the [101]* direction is normal to the (101) surface.}
    \label{fig:orient}
\end{figure}

The tetragonal, rutile crystal structure of RuO$_2$ is depicted in Fig.~\ref{fig:orient}. In Figs.~\ref{fig:c-awiggles} \& \ref{fig:temps} we refer to \textbf{H} oriented along the [101]* orientation. This asterisk notation denotes the $ac$ diagonal in reciprocal space, which is the same direction as the vector normal to the (101) surface in real space~\cite{buerger1937*}.

\section{Density functional theory calculations}
\label{apx_dft}

DFT calculations for RuO$_2$ were performed within the full electron, linearized augmented plane-wave package Wien2K~\cite{wien2k}. The electronic structure was converged on an $11\times  11 \times 16$ Monkhorst-Pack $k$-mesh within the Brillouin Zone of the primitive unit cell using the Generalized Gradient Approximation exchange-correlation potential. The effects of spin-orbit coupling were considered where the easy axis was assumed to be along the [001] direction. RLO states were added to the basis for Ru to improve the modeling of its $p$-semicore states. No RLOs were added for O. 

We assumed that P4/\textit{mnm} RuO$_2$ adopts lattice parameters 4.496799768~\AA, 4.496799768~\AA,  3.1049001876~\AA. Within the unit cell there are two equivalent Ru sites at X=0, Y=0, Z=0 and X=0.5, Y=0.5, Z=0.5 and four equivalent oxygen sites: 0.30530000 Y=0.30530000 Z=0.00000000, X=0.69470000 Y=0.69470000 Z=0.00000000, X=0.19470000 Y=0.80530000 Z=0.50000000, X=0.80530000 Y=0.19470000 Z=0.50000000. 

For frequency analysis of the Fermi surface, we performed a single non-self-consistent DFT iteration of our converged electronic structure, projected onto a $19\times  19 \times 27$ Monkhorst-Pack $k$-mesh. The angular evolution of quantum oscillatory frequencies was determined using SKEAF~\cite{Rourke_2012}, and Fermi surface visualization was performed using py\_FS~\cite{QOscode2}.

\clearpage

\bibliography{RuO2}
\end{document}